\documentclass[aps,prx,reprint,amsmath,amssymb,nolongbibliography]{revtex4-2}
\usepackage[bookmarks=false,linkcolor=NavyBlue,urlcolor=NavyBlue,colorlinks,citecolor=NavyBlue]{hyperref}
\usepackage{graphicx}
\usepackage{amsmath}
\usepackage[svgnames]{xcolor}
\usepackage{enumerate}
\usepackage[normalem]{ulem}
\usepackage{microtype}
\usepackage{comment}

\newcommand{\figref}[2]{\hyperref[#1]{\ref{#1}(#2)}}
\newcommand{\figsref}[2]{\hyperref[#1]{\ref{#1}#2}}
\newcommand{\Weizmann}{Department of Condensed Matter Physics, Weizmann Institute of Science, Rehovot, Israel 7610001}

\begin{document}

\title{Josephson junction arrays as a platform for topological phases of matter}

\author{Omri Lesser}
\affiliation{\Weizmann}

\author{Ady Stern}
\affiliation{\Weizmann}

\author{Yuval Oreg}
\affiliation{\Weizmann}

\begin{abstract}
Two-dimensional arrays of superconductors separated by normal metallic regions exhibit rich phenomenology and a high degree of controllability. We establish such systems as platforms for topological phases of matter, and in particular chiral topological superconductivity. We propose and theoretically analyze several minimal models for this chiral phase based on commonly available superconductor-semiconductor heterostructures.
The topological transitions can be adjusted using a time-reversal-symmetry breaking knob, which can be activated by controlling the phases in the islands, introducing flux through the system, or applying an in-plane exchange field.
We demonstrate transport signatures of the chiral topological phase that are unlikely to be mimicked by local non-topological effects. The flexibility and tunability of our platforms, along with the clear-cut experimental fingerprints, make for a viable playground for exploring chiral superconductivity in two dimensions.
\end{abstract}
\maketitle


\section{Introduction}
Junctions between two superconductors separated by a normal region exhibit rich physics. These so-called superconductor-normal-superconductor (SNS) Josephson junctions (JJs) allow dissipationless current to flow through the normal region, whose direction is determined by the phase difference between the two superconductors~\cite{tinkham_introduction_2004}. In the last decades, this fundamental unit has been extended to two-dimensional (2D) \emph{arrays} of JJs~\cite{fazio_quantum_2001,newrock_two-dimensional_2000}.
Among the myriad of phenomena observed in JJ arrays are the superconductor-insulator transition~\cite{goldman_superconductor-insulator_2010,eley_approaching_2012,bottcher_superconducting_2018}, achieved by varying the ratio between the Josephson coupling between islands to the charging energy of the islands, and the Berezinskii-Kosterlitz-Thouless transition driven by proliferation of vortices which turn the superconductor to a normal metal~\cite{jos_40_2013,bottcher_berezinskii-kosterlitz-thouless_2022}.

Meanwhile, and in a seemingly unrelated trajectory, topology has become one of the prevalent themes in modern condensed matter physics~\cite{hasan_colloquium_2010}. The prime example is the quantum Hall effect, whose remarkable experimental features are extremely robust to imperfections~\cite{girvin_quantum_1999}. This robustness originates from the chiral edge modes that are protected by the topology of the bulk. The superconducting analog of this state is the chiral $p$-wave superconductor, with the canonical model being the $p_{x}+ip_{y}$ superconductor~\cite{alicea_new_2012,bernevig_topological_2013,sato_topological_2017,claassen_universal_2019}. The universal, model-independent features of this topological phase are the appearance of chiral Majorana edge modes and the existence of Majorana bound states in vortex cores. These two features establish two-dimensional $p$-wave superconductors as an intriguing novel state of matter.

There have been several reports of possible chiral topological superconductivity in naturally occuring materials~\cite{wang_evidence_2020,jiao_chiral_2020,nayak_evidence_2021}. However, a clear-cut identification of the topological nature requires tunability --- the ability to switch the topological phase on and off --- which is lacking in these materials.
Much effort has therefore been devoted to engineered platforms. One of the pioneering proposals introduced by Fu and Kane~\cite{fu_superconducting_2008} relied on inducing the superconducting proximity effect to the surface of a three-dimensional topological insulator. This remarkable idea poses two challenges: it is not tunable, and it is hard to realize experimentally~\cite{charpentier_induced_2017}. It was later suggested to use more conventional materials~\cite{sau_generic_2010,alicea_majorana_2010}, but these also come with their experimental complications due to either vortices or material restrictions. 
Several other intriguing ideas have also been put forward in recent years~\cite{qi_chiral_2010,li_two-dimensional_2016,lee_inducing_2017,wang_multiple_2018,lee_topological_2019,rachel_quantized_2017,palacio-morales_atomic-scale_2019,jian_chiral_2021}.

Our goal in this manuscript is to lay the groundwork for combining the physics of Josephson junction arrays with topological matter~\cite{shabani_two-dimensional_2016}.
We propose realistic and viable platforms that give rise to two-dimensional topological superconductivity, in arrays of superconducting islands separated by normal conducting regions with spin-orbit coupling. 

To demonstrate the opportunities available in this platform, we analyze several minimal models. We supplement the heuristic arguments with numerical tight-binding simulations, establishing the existence of superconducting gapped chiral phases.
We introduce three possible ways to form topological superconductivity: phase control (Sec.~\ref{sec:phase-control}), where we find in analogy to previous analysis in one dimension (1D)~\cite{van_heck_single_2014,lesser_three-phase_2021,lesser_phase-induced_2021,lesser_one-dimensional_2022,lesser_majorana_2022} that phase winding is essential; orbital magnetic fluxes (Sec.~\ref{sec:flux}); and \emph{in-plane} exchange field (Sec.~\ref{sec:exchange}).
While the experimental implementation of the three proposals we discuss here may differ significantly, they are all intimately related. In the first proposal, we exert external control over the phases of the superconductor. In the second proposal, the application of a magnetic field induces phase differences between the superconducting islands by creating vortices. Finally, the presence of spin-orbit coupling in the third proposal allows us to view the application of a Zeeman (or exchange) field as an effective orbital vector potential.
We conclude by demonstrating simple and practical experimental signatures of the topological superconducting phases.


\section{Phase-controlled Josephson junction arrays}\label{sec:phase-control}

\begin{figure*}[ht]
    \centering
    \includegraphics[width=\linewidth]{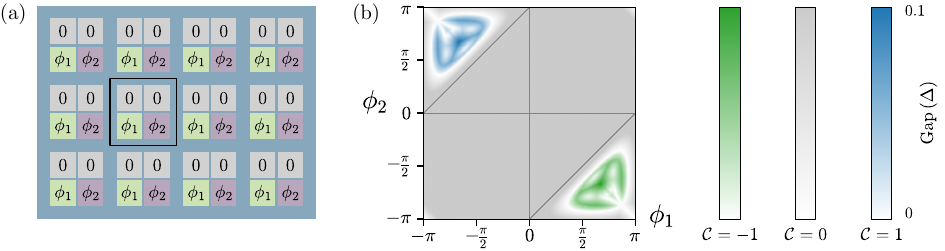}
    \caption{Chiral topological superconductivity in a phase-controlled Josephson junction array.
    (a)~The proposed configuration, consisting of a spin-orbit coupled 2DEG (blue) covered by superconducting islands (gray). The islands form a superlattice with the repeating superconducting phases $\left\{0,0,\phi_1,\phi_2\right\}$; the unit cell is shown in a black frame.
    (b)~Topological phase diagram of the model as a function of the phases $\phi_1$ and $\phi_2$. Colors indicate the Chern number ${\cal C}$, whereas the intensity corresponds to the bulk gap relative to the pairing gap $\Delta$. The topologically non-trivial states with ${\cal C}=\pm1$ appear only in regions where phase winding occurs, which are marked by solid lines for clarity. Notice that the chirality (the sign of the Chern number) flips when changing the phases from forming a vortex to an anti-vortex in each unit cell. In the tight-binding model, each island is represented by a single site, and the parameters are $t=1$, $\Delta = 0.5$, $\mu = 1.15$, $\eta = 0.5$, $\lambda = 0.2\pi$
    \label{fig:phases_sq}}
\end{figure*}

A useful starting point for the considerations we apply is inspired by the formation of topological superconductivity on the surface of a three-dimensional topological insulator~\cite{fu_superconducting_2008,qi_topological_2011}.
The presence of a single Fermi surface forces the superconducting state to be topological, such that each vortex core carries one Majorana mode at zero energy, and the superconducting state is encircled by a chiral edge mode. In the system we consider, the normal state is a stand-alone two-dimensional system in which there are two Fermi surfaces, and the spin-orbit coupling makes the radii of these surfaces unequal. When superconductivity is induced in such a system without coupling the two Fermi surfaces, two copies of a topological superconductor will be formed with opposite chiralities. Then, each vortex will carry \emph{two} Majorana modes. An array of vortices will then form a tight-binding array of sites, with each sites carrying two Majorana modes. The interaction between the normal and superconducting regions will then couple the Majorana modes with intra- and inter-site tunneling. When tuned properly, this coupling will create a 2D topological superconductor.

We now construct a model for a topological superconductor that is based on these premises. We consider a 2D electron gas (2DEG) with Rashba spin-orbit coupling.
We then selectively introduce superconducting islands arranged in a square lattice, as shown in Fig.~\figref{fig:phases_sq}{a}, and assume full control of the phase of each island and the Josephson coupling of nearest-neighboring islands. The islands form a periodic structure, with a supercell which we take to have four islands. 

Based on the considerations above, as well as on previous results in related models~\cite{fu_superconducting_2008,van_heck_single_2014,lesser_one-dimensional_2022}, we seek a phase configuration that includes a vortex, i.e., a phase winding of $2\pi$, within each cell. The periodicity of the phases implies that when properly defined, the net phase winding over a unit cell must vanish, such that each unit cell would include both a vortex and an anti-vortex. Furthermore, our choice of phase configuration should avoid a symmetry that forbids a superconductor with a non-zero Chern number and chiral edge states. Such symmetries include time-reversal symmetry, mirror symmetries, and combinations of these symmetries with partial unit cell translations. 

With these considerations in mind, we assign the four islands of the unit cell the phases of of $\left\{0,0,\phi_1,\phi_2\right\}$, and choose the intra-cell Josephson couplings to be larger than the inter-cell ones. A proper choice of $\phi_1,\phi_2$ breaks time-reversal symmetry and introduces phase windings into the unit cells, while the variation of the Josephson couplings breaks mirror symmetries. In practice, the strength of the Josephson coupling is to be controlled by the width of the normal regions between the superconducting islands. 

We model the system using the tight-binding method on a square lattice. 
The topological invariant characterizing this 2D class D system is the integer-valued Chern number~${\cal C}$, which counts the number of chiral Majorana edge modes~\cite{altland_nonstandard_1997,schnyder_classification_2008,ryu_interacting_2012,kitaev_periodic_2009}. The number ${\cal C}$ may be evaluated explicitly in momentum space~\cite{fukui_chern_2005,lisle_detection_2014}. It is also informative, and much easier, to calculate its parity,
\begin{equation}\label{eq:pfaffian}
    \left(-1\right)^{\cal C} = \prod_{\vec{k}\in{\rm TRIM}} {\rm Pf}\left[\Lambda {\cal H}(\vec{k}) \right],
\end{equation}
where TRIM are the time-reversal-invariant momenta, ${\cal H}$~is the Bogoliubov--de Gennes Hamiltonian, $\Lambda$ is the anti-unitary particle-hole operator, and ${\rm Pf}(\cdot)$ is the Pfaffian.

\begin{figure*}[t]
    \centering
    \includegraphics[width=\linewidth]{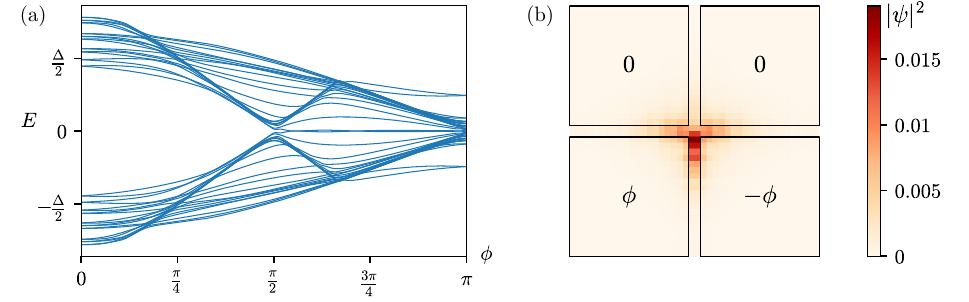}
    \caption{Four phase-biased islands.
    (a)~Energy spectrum (lowest 40 levels) as a function of the phase bias $\phi$, showing many Andreev bound states. When the phase winds ($\phi>\pi/2$), the lowest Andreev bound states are close to zero energy.
    (b)~Wavefunction of the lowest-energy state at $\phi=3\pi/4$, showing anisotropic localization at the junction between the four islands.
    The tight-binding parameters are $t=1$, $\lambda=\pi/5$, $\mu=-3$, $\Delta=0.3$.
    \label{fig:four_islands}}
\end{figure*}

Within a tight-binding model, the normal state, which  is a 2DEG with Rashba spin-orbit coupling, is described by the following discrete Hamiltonian,
\begin{equation}
\begin{aligned}H_{0}= & \sum_{m,n;s,s'}-\mu_{m,n}c_{m,n,s}^{\dagger}c_{m,n,s}+\\
 & \left[-c_{m,n,s}^{\dagger}\left(t^{x}_{m,n}e^{i\lambda\sigma_{y}}\right)_{ss'}c_{m+1,n,s'}\right.\\
 & \left.-c_{m,n,s}^{\dagger}\left(t^{y}_{m,n}e^{-i\lambda\sigma_{x}}\right)_{ss'}c_{m,n+1,s'}+\text{H.c.}\right],
\end{aligned}
\end{equation}
where $c,c^{\dagger}$ are electronic creation and annihilation operators, the indices $m,n$ label the sites in the $x,y$ directions, $s,s'\in\left\{\uparrow,\downarrow\right\}$ are the spin indices, $\sigma$ are Pauli matrices in spin space, $\mu_{m,n}$ is the (site-dependent) chemical potential, $t$ is the nearest-neighbor hopping amplitude, and $\lambda$ is the spin rotation angle due to Rashba spin-orbit coupling. The hopping is modulated, to generate the varying strength of the Josephson coupling: within the unit cell it is $t$, and between unit cells it is $\eta t$ where $\eta<1$. The full Hamiltonian is given by $H=H_{0}+H_{\rm SC}$. Here the superconducting part is 
\begin{equation}
    H_{\text{SC}}=\Delta\sum_{m,n}e^{i\phi_{m,n}}c_{m,n,\uparrow}^{\dagger}c_{m,n,\downarrow}^{\dagger}+\text{H.c.},
\end{equation}
where $\Delta$ is the pairing potential and the site-dependent phase configuration $\phi_{m,n}$ is illustrated in Fig.~\figref{fig:phases_sq}{a}. We use the momentum-space representation to calculate topological invariants and energy gaps. This Hamiltonian is somewhat simplified in having the unit cell for the superconducting islands of the same size as the unit cell of the normal part. However, we do not expect this simplification to affect the analysis of the topological properties of the system, which are our main focus here. 

Remarkably, by varying $\phi_1$ and $\phi_2$, we observe two distinct topologically non-trivial phases, for which ${\cal C}=\pm1$, as shown in Fig.~\figref{fig:phases_sq}{b}. These phases differ in their opposite chirality. They possess an excitation gap which is a significant fraction of the induced pairing gap $\Delta$. In practice, we expect this gap to depend on many parameters of the system, such as the widths of the normal regions separating the superconducting islands.

\begin{figure*}
    \centering
    \includegraphics[width=\linewidth]{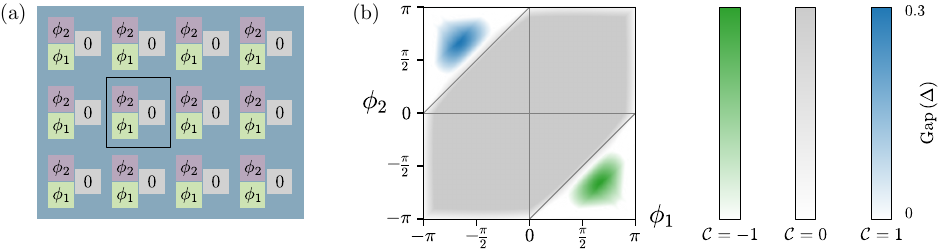}
    \caption{2D topological superconductivity in a phase-controlled Josephson junction array where each unit cell contains three islands.
    (a)~The proposed geometry: 2DEG (blue) with superconducting islands (gray, green, and purple, with the color coding representing the superconducting phases) on top. The unit cell is depicted in a black frame.
    (b)~Topological phase diagram of the model as a function of the phases $\phi_1$ and $\phi_2$. Colors indicate the Chern number ${\cal C}$, whereas the intensity corresponds to the bulk gap relative to the pairing gap $\Delta$. As in the model of Fig.~\ref{fig:phases_sq}, the topologically non-trivial states with ${\cal C}=\pm1$ appear only in regions where phase winding occurs, which are marked by solid lines for clarity.
    The tight-binding parameters in the simulation are $t=1$, $t'=0.5$, $\mu=-1$, $\Delta=0.5$, $\lambda=0.3\pi$, and the two left islands in each unit cell form an angle $\theta=0.3\pi$ with respect to the right island.
    \label{fig:phases_tri}}
\end{figure*}

To facilitate understanding the subgap spectrum, we begin by considering a single unit cell, and model the four islands within it as many sites ($20\times 20$) separated by a few sites (2 in each direction) without superconductivity. The entire system is deposited on a spin-orbit coupled medium. 
The energy spectrum of the four-island system is shown in Fig.~\figref{fig:four_islands}{a}. At zero phase bias, we observe Andreev bound states starting around the energy $\Delta/2$. These states are merely a result of the finite normal region. As $\phi$ grows, the Andreev bound states penetrate deeper into the gap, until at $\phi=\pi/2$ --- the onset of phase winding --- the lowest state reaches zero energy. We then observe several states at low energies that can be thought of as the discrete version of the Caroli-de Gennes-Matricon states appearing in vortex cores. Remarkably, for a range of phase differences between $\pi/2$ and $\pi$, the vortex carries a pair of states close to zero energy, whose distance from the other states is rather large (a significant fraction of $\Delta$). Then, the low-energy spectrum of the array may be described in terms of two Majorana bands, which originate from the coupling of these two states in nearest-neighboring vortices. Notice that the  lowest states are in no way topologically protected. At $\phi=\pi$, we effectively get a single $\pi$ Josephson junction between the upper and lower islands, resulting in many states close to zero energy. Figure~\figref{fig:four_islands}{b} shows the wavefunction (local density) of the lowest-energy state at $\phi=3\pi/4$ (i.e., with phase winding). This state is localized at the junction between the four islands, with a larger weight at the junction between the $\phi$, $-\phi$ islands.

The particular phase configuration we presented, described by $\phi_1$ and $\phi_2$, is just an example for a broader principle, namely the construction of a two-dimensional topological superconductor with a non-zero Chern number by the combination of a periodic, vortex-carrying phase pattern, with Rashba spin-orbit coupling. 
As expected, the Chern number switches sign when both phases switch signs ($\phi_1\to-\phi_1$, $\phi_2\to-\phi_2$), corresponding to time reversal, or when the two phases are exchanged, $\phi_1\to\phi_2$, $\phi_2\to\phi_1$, corresponding to a mirror operation. 
These symmetries constrain the shape of the phase diagram, as evident also in the numerical results shown in Fig.~\figref{fig:phases_sq}{b}.
However, our setup is by no means limited to just this type of configuration.

As an example, we consider a three-island unit cell, as depicted in Fig.~\figref{fig:phases_tri}{a}. As in the model analyzed in the main text, the Josephson coupling within each unit cell, $t$, is larger than that between unit cells, $t'$. 
Figure~\figref{fig:phases_tri}{b} shows the topological phase diagram of the model. We find very similar features to the phase diagram of Fig.~\figref{fig:phases_sq}{b}: the chiral states appear only when the three phases wind. The same symmetries apply to this diagram; in particular, flipping the phases must flip the Chern number. 

\begin{figure}[b]
    \centering
    \includegraphics[width=\linewidth]{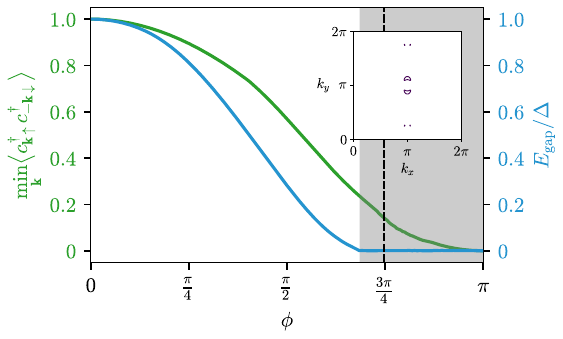}
    \caption{Gapless superconductivity in the phase-controlled Josephson junction array of Fig.~\ref{fig:phases_sq}. Setting $\phi_1=-\phi_2=\phi$, we calculated the single-particle energy gap (blue) and the pair correlation function (green). For both quantities, we looked for the lowest values across the two-dimensional Brillouin zone. At a certain region (shaded), the single-particle gap vanishes while the pair correlation function remains finite at all momenta. This is the characteristic behavior of a gapless superconductor.
    Inset: $E=0$ Fermi surface at $\phi=3\pi/4$ (marked by a dash line) inside the gapless phase.
    \label{fig:gapless}}
\end{figure}

We notice that certain regions in the phase diagram in Fig.~\figref{fig:phases_sq}{b} are gapless. The appearance of gapless lines in the transitions between phase with different Chern numbers is well understood, but the seemingly extended regions in parameter space where gapless phases emerge are not \emph{a priori} expected. Upon closer inspection, we find that these regions have a \emph{finite Fermi surface} at zero energy, implying that they are not nodal superconductors (in which the gap closes only at an isolated point in the Brillouin zone). As we demonstrate in Fig.~\ref{fig:gapless}, this behavior is consistent with \emph{gapless superconductivity}, where the single-particle energy gap vanishes but the pair correlations remain non-zero~\cite{maki_gapless_1969}. Such a situation is known to occur in superconductors with many magnetic impurities or large current, and our platform provides another setup to study this phase.

To understand the origin of the gapless phase in our system, it is useful to start from the point of view of Majorana bound states in vortex cores, as in Fig.~\ref{fig:four_islands}. When a vortex is present, the low-energy spectrum consists of two coupled Majorana bound states, whose energies are slightly shifted from zero. Our system can therefore be modeled as a square lattice with two states per site, making it a two-band model. The only symmetry present is particle-hole, which in real space we can take to be $\tau_{x}{\cal K}$, where $\tau_j$ are the Pauli matrices acting in the $2\times2$ band space and $\cal K$ refers to complex conjugation. The most general Hamiltonian allowed by this symmetry, assuming only nearest-neighbor hopping, is 
\begin{equation}
\begin{aligned}H & =\left(a_{0}\sin k_{x}+b_{0}\sin k_{y}\right)\tau_{0}+\left(a_{1}\sin k_{x}+b_{1}\sin k_{y}\right)\tau_{x}\\
 & +\left(a_{2}\sin k_{x}+b_{2}\sin k_{y}\right)\tau_{y}+\left(a_{3}\sin k_{x}+b_{3}\sin k_{y}+c\right)\tau_{z}
\end{aligned}
\end{equation}
where the coefficients $a_j,b_j,c$ are real. The corresponding spectrum is
\begin{equation}
\begin{aligned}E & (k_x,k_y) =a_{0}\sin k_{x}+b_{0}\sin k_{y}\\
 & \pm\left[\left(a_{1}\sin k_{x}+b_{1}\sin k_{y}\right)^{2}+\left(a_{2}\sin k_{x}+b_{2}\sin k_{y}\right)^{2}\right.\\
 & \left.+\left(a_{3}\sin k_{x}+b_{3}\sin k_{y}+c\right)^{2}\right]^{1/2}.
\end{aligned}
\end{equation}
To find a gapless phase, we need to solve $E(k_x,k_y)=0$; this is a single equation with two variables, $k_x$ and $k_y$, and in general the solution is a curve. In the absence of $a_0$ and $b_0$, finding a solution would require all terms in the square root to vanish at the same time (since they are all non-negative). This would impose many constraints on the coefficients and is thus not expected to have a solution generically. However, having $a_0,b_0\neq0$ (which happens naturally for systems with a superconducting phase gradient) relaxes this strict condition and makes the gapless phase generic if all coefficients are of the same order of magnitude. We note that if, for example, the Majoranas are very far away from zero energy, as in the case without a vortex, $c$ is much larger than the other coefficients and generically we will \emph{not} find a gapless phase.


\section{Vortex lattice formed by flux}\label{sec:flux}

\begin{figure*}[t]
    \centering
    \includegraphics[width=\linewidth]{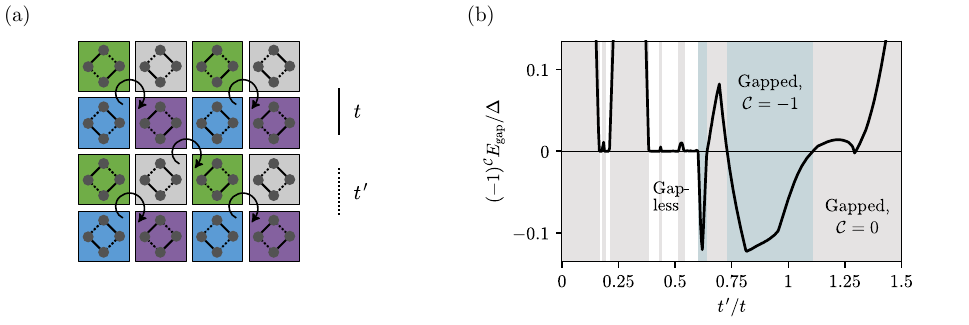}
    \caption{Chiral topological superconductivity in a Josephson junction array with applied magnetic flux.
    (a)~Proposed configuration, where each island (square) is described by four sites (circles). The islands are connected to their neighbors by Josephson coupling $t_{\rm J}$, and the intra-island tunneling is either $t$ (solid lines) or $t'$ (dashed lines). The islands' phases (depicted by green, gray, blue, or purple shading) are chosen in accordance with the checkerboard vortex lattice appropriate for $h/4e$ flux per plaquette (vortices are shown as winding arrows).
    (b)~Topological phase diagram as a function of $t/t'$, featuring gapped topological (blue), gapped non-topological (gray), and gapless (white) phases. The diagram shows the parity of the Chern number ${\cal C}$ (which is $-1$ for the topologically non-trivial phases), multiplied by the bulk energy gap in units of the pairing gap $\Delta$.
    The tight-binding parameters $t=1$, $\Delta = 0.1$, $\mu = 0.42$, $\lambda = 0.4\pi$, and $t_{\rm J}=t$.
    \label{fig:flux}}
\end{figure*}

Motivated by the results of the previous section, we now ask whether a chiral topological state can be induced without controlling the phase of each island directly. Rather, the phases are to be controlled by subjecting the sample to an out-of-plane magnetic flux, such that a vortex lattice is formed~\cite{halsey_josephson-junction_1985,lankhorst_annealed_2018}. Since the basic building block we used in the previous section was a unit cell with phase winding, it is reasonable to expect a similar mechanism to work when flux is applied. 

We demonstrate this concept using the special case of $f=1/2$ superconducting flux quanta per plaquette, in an array in which the London length is much larger than the unit cell. In such an array the magnetic field is approximately uniform, with a flux of $h/4e$ per plaquette, but the vortices form a vortex lattice with a doubled unit cell, breaking the translation symmetry of the original lattice. Every second plaquette hosts a vortex, and the vortices form a checkerboard lattice~\cite{halsey_josephson-junction_1985,lankhorst_annealed_2018}. Taking into account the frustration of some of the bonds due to the vector potential, one can extract the superconducting phases corresponding to this vortex configuration. As in Sec.~\ref{sec:phase-control}, we place the islands on top of a spin-orbit coupled 2DEG.

In the proposed configuration, the creation of a chiral topological superconductor again requires all mirror symmetries to be broken, such that vortices are distinguished from anti-vortices. 
We again do that by creating a distinction between the unit cells and their surrounding, as illustrated in Fig.~\figref{fig:flux}{a}: each island is split into four sites, and the tunneling between them are not homogeneous.

In the tight-binding simulation, the hopping is modulated as depicted in Fig.~\figref{fig:flux}{a}. In addition, to take into account the effect of the magnetic field on the electrons, we use the Peierls substitution~\cite{peierls_zur_1933}, which makes the hopping along the $x$ direction complex (we use the gauge $\vec{A}=By\hat{x}$):
\begin{equation}
\begin{aligned}
    & c_{m,n,s}^{\dagger}\left(te^{i\lambda\sigma_{y}}\right)_{ss'}c_{m+1,n,s'}\\
    \to & -c_{m,n,s}^{\dagger}\left(te^{i\frac{\pi f}{2}n}e^{i\lambda\sigma_{y}}\right)_{ss'}c_{m+1,n,s'}.
\end{aligned}
\end{equation}
The flux also induces a non-uniform phase configuration, which in our gauge is $\left\{ 0, 7\pi/4, \pi/4, \pi/2 \right\}$ with the order being \{blue, gray, purple, green\} in the notation of Fig.~\figref{fig:flux}{a}.

By tuning the tunneling amplitudes between the sites in each island, we observe gapped topological, gapped non-topological, and gapless phases [see Fig.~\figref{fig:flux}{b}]. We find a 2D topological phase whose gap is an appreciable fraction of the induced pairing gap $\Delta$. Again, in practice, we expect this gap to depend on many parameters of the system, such as the widths of the normal regions separating the superconducting islands.

We notice that the unit cell required to model this minimal system consists of eight islands. That is because the Aharonov-Bohm flux per plaquette is $1/4$ in units of $h/e$ (the non-superconducting flux quantum), and therefore the magnetic unit cell describing the normal tunneling is composed of four plaquettes. The overall topological phase results from the breaking of time-reversal symmetry both by this Aharonov-Bohm effect and by the superconducting phase winding. The theoretical analysis becomes slightly more complicated due to the orbital effect, but on the other hand, the experimental realization is a bit simpler compared to the individual phase control scheme proposed in Sec.~\ref{sec:phase-control}.

The possibility of controlling the phases in a JJ array, either directly or by external magnetic flux, opens up another research avenue related to topology. We envision a system with alternating phase gradients, such that the phase gradient is $\pi g_1$ in even rows and $\pi g_2$ in odd rows. If $g_1\neq g_2$, the phase pattern becomes periodic if $g_1$ and $g_2$ are rational. If either $g_1$ or $g_2$ is irrational, then the phase pattern is aperiodic, and the vortices form a quasicrystal. Furthermore, if the conditions of a topological state are met, then each vortex hosts a Majorana zero modes and we get a 2D quasicrystal of Majoranas. 
Such a setup can be thought of as a superconducting analog of the Hofstadter butterfly~\cite{hofstadter_energy_1976,aidelsburger_artificial_2016,lesser_universal_2020} without needing large magnetic fields. 
The unique features related to Hofstadter physics are rarely observed in experiment~\cite{geisler_detection_2004,hunt_massive_2013,dean_hofstadters_2013,ponomarenko_cloning_2013,aidelsburger_artificial_2016,lu_multiple_2021}, thus the possibility of finding them in a rather simple setup highlights the various opportunities offered by JJ array.


\section{Island arrays subjected to an in-plane exchange field}\label{sec:exchange}

There is another valuable experimental knob that is accessible in JJ arrays and that we have not used so far, and that is an in-plane exchange field. Such a field can be induced by applying an external magnetic field or by coupling the system to a ferromagnet that induces an exchange field. We notice that for thick samples, external in-plane fields also have orbital effects, which alter the phases of the islands~\cite{banerjee_signatures_2023,banerjee_local_2023}; therefore, it might be preferable to use a ferromagnet such as EuS~\cite{vaitiekenas_zero-bias_2020}. Superconductivity generally survives under much larger fields when they are applied in-plane than out-of-plane~\cite{tinkham_introduction_2004}; this experimental knob allows us to study the unique properties of the time-reversal-broken phase of JJ arrays. In this section, we show that by controlling the geometry of the island array and the orientation of the in-plane field, a chiral topological superconducting phase emerges.

In a Rashba 2DEG uniformly covered by a superconductor, only an out-of-plane magnetic field can open a topological gap~\cite{sau_generic_2010}. The underlying reason for this is that for an in-plane field, one can always find a momentum direction along which the field-free Hamiltonian commutes with the Zeeman term, and therefore a gap will not open. In other words, the Brillouin zone contains a point that will not gap out. Here we argue that JJ arrays provide a natural way to overcome this problem: the periodic modulation they correspond to is able to mix this point with higher-momentum states and thus gap it out. 

To illustrate our approach, consider first a highly anisotropic system comprising an array of long superconducting wires with spin-orbit coupling. When applying a magnetic field along the wires' axis and tuning the chemical potential properly, they all become 1D topological superconductors, with Majorana zero modes at their edges~\cite{lutchyn_majorana_2010,oreg_helical_2010}. We choose the lattice layout such that these zero modes form a decorated honeycomb lattice~\cite{yao_exact_2007}. If we then allow tunneling between the wires, the effective model is a Majorana honeycomb with next-nearest-neighbor hopping. This model is well studied in the context of Kitaev's model for a $\mathbb{Z}_2$ spin liquid, and in the superconducting case, it corresponds to a gapped topological phase~\cite{kitaev_anyons_2006}.

We note that platforms consisting of adatoms on the surface of a superconductor, arranged in the same decorated honeycomb form proposed here, may also be utilized to create two-dimensional topological superconductivity. The array we propose here is sparse and therefore expected to have less strain effects compared to a dense array of adatoms. Recent experiments employed atomic scanning tunneling microscopy (STM) manipulation to arrange ferromagnetic chains, showing signatures of topological phases~\cite{ mier_atomic_2021} and arrays of adatoms with topological properties have recently been fabricated~\cite{soldini_two-dimensional_2023}.

\begin{figure*}[t!]
    \centering
    \includegraphics[width=\linewidth]{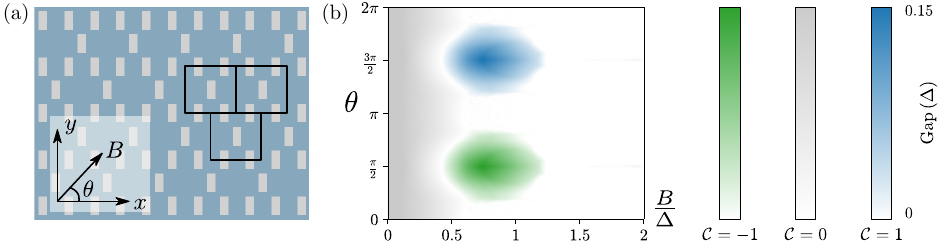}
    \caption{Chiral topological superconductivity in a Josephson junction array with an in-plane exchange field.
    (a)~The proposed configuration, with superconducting islands (gray) deposited on top of a Rashba 2DEG (blue). Several unit cells are indicated in frames, and they repeat themselves in a brick-wall lattice. An in-plane exchange field $B$ is applied at an angle $\theta$ with respect to the horizontal axis $\hat{x}$.
    (b)~Topological phase diagram as a function of $B$ and $\theta$. Colors indicate the Chern number ${\cal C}$, whereas the intensity corresponds to the bulk gap relative to the pairing gap $\Delta$. We find that $\theta=\frac{\pi}{2},\frac{3\pi}{2}$, i.e., along the elongated islands~($\hat{y}$), are most favorable for a chiral topological state (${\cal C}=\pm1$) with a large gap.
    In the tight-binding simulation, we modeled islands that are twice as long as they are wide, and used the parameters $t=1$, $\mu_{\rm S} = -1.2$ (chemical potential in the superconducting regions), $\mu_{\rm N} = 3.3$ (chemical potential in the normal regions), $\Delta=0.2$, $\lambda = 0.2\pi$.
    \label{fig:exchange}}
\end{figure*}

We now make the model more realistic by replacing the wires with superconducting islands deposited on top of a Rashba 2DEG; see Fig.~\figref{fig:exchange}{a}. The size of the islands and the spacings between them are all design parameters at our disposal. To make contact with the intuitive wire-based picture, we consider slightly elongated islands; however, we have verified that square islands yield very similar results. The unit cell repeats itself in a triangular lattice.

We simulate the system using the same tight-binding methods described in Sec.~\ref{sec:phase-control} and add an exchange term to the Hamiltonian, 
\begin{equation}
    H_{\text{ex}}=B\sum_{m,n;s,s'}c_{m,n,s}^{\dagger}\left(\sigma_{x}\cos\theta+\sigma_{y}\sin\theta\right)_{ss'}c_{m,n,s'}.
\end{equation}
We model the superconducting structure as 
\begin{equation}
    H_{\text{SC}}=\sum_{m,n}\Delta_{m,n}c_{m,n,\uparrow}^{\dagger}c_{m,n,\downarrow}^{\dagger}+\text{H.c.},
\end{equation}
where $\Delta_{m,n}=\Delta$ in the islands depicted in Fig.~\figref{fig:exchange}{a} and zero elsewhere.
We find gapped topological regions in parameter space [see Fig.~\ref{fig:exchange}{b}]. In this configuration, we find that the preferred direction of the magnetic field is $\hat{y}$ (notice that the $\hat{x}$ and $\hat{y}$ directions are not equivalent due to the triangular lattice). This can be seen as a remnant of the toy-model construction, where the system is most robust when the field is applied along the 1D wires. As in the phase-controlled setup of Sec.~\ref{sec:phase-control}, we find extended regions of gapless superconductivity in the phase diagram.


\section{Experimental signatures}\label{sec:transport}

\begin{figure*}[t]
    \centering
    \includegraphics[width=\linewidth]{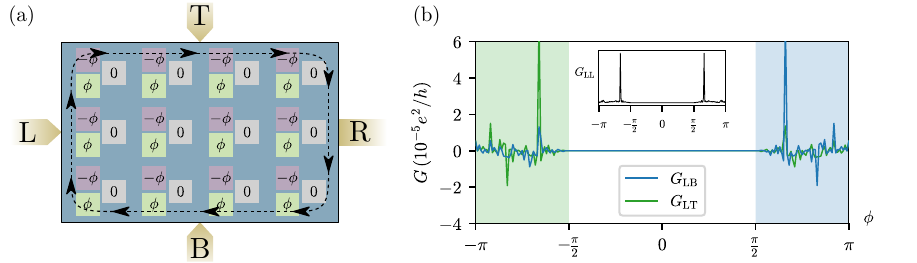}
    \caption{
    Transport signatures of chiral topological superconductivity.
    (a)~Proposed measurement setup: a variant of the phase-controlled island array of Sec.~\ref{sec:phase-control} is contacted by three tunnel probes (L, B, T) and one strongly coupled lead (R). We envision the L lead as the source and the rest as drains. The chiral edge mode appearing in the topological phase is illustrated in dashed lines.
    (b)~Transport simulation results at zero bias voltage as a function of $\phi$. The non-local conductances $G_{\rm LB}$ and $G_{\rm LT}$ onset only in the topological phases, and they appear for Chern numbers ${\cal C}=+1,-1$ respectively, due to the chiral nature of the edge mode. The inset shows the local conductance $G_{\rm LL}$, which also onsets only in the topological phase (the ${\cal C}=0$ phase is fully gapped), but does not differentiate between the chiralities ${\cal C}=\pm 1$. The tight-binding parameters are the same as in Fig.~\ref{fig:phases_tri}, and we use a lattice of $300\times300$ unit cells. The left, top, and bottom leads are four sites wide, whereas the right lead is ten sites wide. 
    \label{fig:transport}}
\end{figure*}

We now address question of how one can tell whether a superconductor under investigation is topological. Following early ideas~\cite{stern_proposed_2006,bolech_observing_2007,akhmerov_electrically_2009,fu_probing_2009}, we focus on transport experiments and propose a simple setup for detecting chiral superconductivity. We note, before going to details, that the existence of an edge mode that surrounds the system makes the identification of 2D topological superconductivity easier in principle than that of its 1D counterpart. 

The measurement setup we consider is shown in Fig.~\figref{fig:transport}{a}. The 2D topological superconductor is connected to four leads, at the left (L), right (R), top (T), and bottom (B). We assume that leads L, T, and B are tunnel coupled to the system, whereas lead R is strongly coupled to the system. In the non-topological phase, single electrons cannot tunnel into the system, and therefore we expect the low-energy conductance to be very low. In the topological phases, on the other hand, current can flow along the edge by virtue of the chiral Majorana mode, and therefore the conductance from L to either T or B (depending on the chirality) will be non-zero. Since lead R is strongly connected to the system, all leftover current is collected there, leading to a very asymmetric chirality-dependent behavior. This phenomenology is indeed seen in our transport simulations~\cite{groth_kwant:_2014}, as demonstrated in Fig.~\figref{fig:transport}{b}.

The simple setup discussed above already showcases the unusual and highly non-local signatures of the topological phase. One can go a step further towards a true interference experiment, for example by coupling the R lead very weakly to the system. Then, the current measured at the T and B leads will correspond to a coherent summation over all possible trajectories leading to them. In this scenario, we expect the currents to be sensitive to the number of vortices present in the bulk: each of these vortices carries a localized zero-energy Majorana bound state, and thus their parity controls the boundary conditions of the chiral Majorana mode that surrounds the system. Inserting a discrete vortex can be done by controlling the phases in the islands or by changing an external magnetic field. Furthermore, in this setup one can also create domain walls, for example by flipping the signs of $\phi_1$ and $\phi_2$ in a certain region of the system, opening the door to interference experiments akin to the idea of Ref.~\cite{akhmerov_electrically_2009}.

Beyond charge transport, heat transport measurements also provide valuable information on the nature of topological systems~\cite{pekola_colloquium_2021}. A prime example is the half-quantized thermal Hall conductivity that was observed in the fractional quantum Hall regime~\cite{banerjee_observation_2018}. In the case of 2D $p$-wave superconductors, we expect a similar phenomenology: the heat transport should be directional, due to the chiral nature of the system, and the thermal conductivity should be quantized, providing a direct measurement of the Chern number~\cite{nomura_detailed_2002,shakeripour_heat_2009,rosenstein_effect_2013,murray_majorana_2015}. We note, however, that such measurements are typically more challenging experimentally than charge transport measurements.

As mentioned above, it is possible to introduce vortices into the system. For a chiral topological superconductor, each vortex host Majorana bound state(s). In principle, these vortices can be moved around: by controlling the phases of the islands, the positions of the Majorana bound states are manipulated. Since these states are predicted to be mutual non-Abelian anyons, this could serve as the starting point of experiments aimed at uncovering non-Abelian exchange statistics.


\section{Conclusion}
Much progress has been made in recent years in the field of topological superconductivity, particularly in quasi-1D systems~\cite{lutchyn_majorana_2018,flensberg_engineered_2021}. In 1D setups, the hallmark of the topological phase is the existence of localized Majorana zero modes at the edges~\cite{mourik_signatures_2012,das_zero-bias_2012,albrecht_exponential_2016,fornieri_evidence_2019,ren_topological_2019,banerjee_signatures_2023,banerjee_local_2023}. Although both theory and experiment are well developed, the distinction of topological effects from non-topological ones in experiments is challenging~\cite{kells_near-zero-energy_2012,vuik_reproducing_2019,pan_physical_2020,hess_local_2021}. In particular, impurities in quasi-1D can easily mimic the signatures of Majorana zero modes, making it hard to agree on the interpretation of experimental data. On the contrary, 2D topological superconductors exhibit \emph{chiral edge modes}, rather than localized states, and are therefore expected to provide much more compelling experimental signatures, especially in non-local conductance.

We have introduced several platforms for 2D chiral topological superconductivity based on arrays of Josephson junctions. We have also shown that straightforward transport measurements can provide unique and robust signatures of these topological phases, unlike zero-bias conductance peaks in localized Majorana zero modes which can be ambiguous. JJ arrays are well developed and have been studied for many years, and therefore we expect our proposal to be within experimental reach. Naturally, more detailed modeling of the particular materials of choice will be needed to provide quantitative predictions. However, as the main requirement for our scheme is strong spin-orbit coupling, several existing and well-established material platforms seem like natural candidates: InAs and InSb quantum wells in proximity with Al~\cite{ke_ballistic_2019,fornieri_evidence_2019,banerjee_signatures_2023}, HgTe quantum wells in proximity with Nb~\cite{reuther_gate-defined_2013,ren_topological_2019}, thin transition metal dichalcogenides layers~\cite{qian_quantum_2014,manzeli_2d_2017} and LaAlO$_3$/SrTiO$_3$ or LaAlO$_3$/KTaO$_3$ interfaces~\cite{cheng_tunable_2016,yu_nanoscale_2022}.

The island arrays we propose are highly tunable, as they offer numerous experimental knobs: gate voltages (densities), phase control, and in-plane exchange field. Furthermore, the island structure has another inherent advantage: it facilitates studying the superconductor-insulator transition~\cite{sondhi_continuous_1997} in a controllable manner~\cite{schon_quantum_1990,bottcher_superconducting_2018,bottcher_berezinskii-kosterlitz-thouless_2022,bottcher_dynamical_2022}. This highly versatile platform will therefore allow one to study the insulating side of the $p_{x}+ip_{y}$ superconductor, which is expected to be very exotic~\cite{terhal_majorana_2012,roy_quantum_2017,wang_gapped_2013,sagi_spin_2019,ebisu_supersymmetry_2019}. Detailed investigation of possible spin-liquid phases in this system is left for a future study.

\begin{acknowledgements}
We thank Charles M. Marcus, Saulius  Vaitiekenas, and Karsten Flensberg for fruitful discussions.
The code used for simulating the models and generating the plots in this study is available at Ref.~\cite{SourceCode}.
This work was supported by the European Union's Horizon 2020 research and innovation programme (Grant Agreement LEGOTOP No. 788715), the DFG (CRC/Transregio 183, EI 519/7-1), ISF Quantum Science and Technology (2074/19), the BSF and NSF (2018643).
\end{acknowledgements}


\appendix

\section{Wire construction}\label{app:wires}
Here we present an additional point of view on constructing a two-dimensional topological superconductor, complementary to the one described in Sec.~\ref{sec:phase-control}. 
This point of view is inspired by the wire construction of an integer quantum Hall state~\cite{yakovenko_quantum_1991,sondhi_sliding_2001}.
It starts from one-dimensional topological superconductivity that is entirely phase-controlled. Schemes for creating such systems were proposed in Refs.~\cite{lesser_three-phase_2021,lesser_phase-induced_2021,lesser_one-dimensional_2022,melo_supercurrent-induced_2019,lesser_majorana_2022}. The key idea is combining a discrete superconducting vortex with a spin-orbit coupled material. 

\begin{figure}
    \centering
    \includegraphics[width=\linewidth]{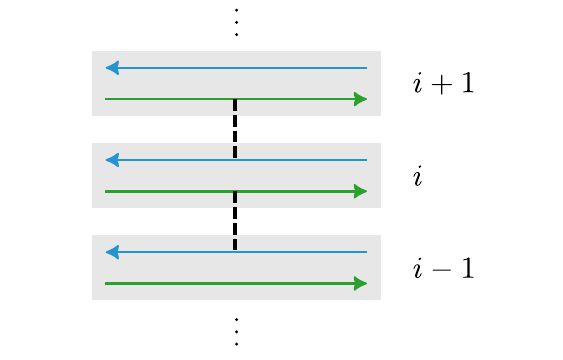}
    \caption{Two-dimensional topological superconductivity from an array of parallel one-dimensional topological superconductors. When each 1D system is tuned to the critical point, it hosts two gapless Majorana modes. Coupling these left- and right-movers (blue and green arrows) appropriately, as shown by the dashed lines, results in a gapless chiral edge mode.
    \label{fig:wire_construction}}
\end{figure}

Concretely, we consider here the quasi-1D model of Ref.~\cite{lesser_one-dimensional_2022}, where the topological phase transition is controlled by two phase differences. A transition between topological and trivial phase may be induced by fine-tuning the parameters of the model (phase differences,  chemical potential, the strength of the spin-orbit coupling, etc.). At the transition point, the gap closes, and the system hosts 
two counter-propagating Majorana modes, a left-mover and a right-mover. When a two-dimensional plane is formed as an array of parallel such one-dimensional systems, the resulting two-dimensional phase depends on the coupling of these systems.  
If the left-mover of the  $i$'s system is only coupled to the right-mover of the $i+1$, then the coupling will gap both modes, leaving at low energy only the chiral modes of the outermost wires. Then, the resulting 2D system is in a chiral topological state~\cite{su_solitons_1979,rice_elementary_1982}. Figure~\ref{fig:wire_construction} visualizes this idea.

We note that, in order to get a chiral topological phase (a 2D gapped phase with a non-zero Chern number), all mirror symmetries must be broken. Our spin-orbit coupled system has the mirror symmetry ${\cal M}_{x}=(x\to-x)\sigma_x$, and therefore a uniform 1D systems are not sufficient for constructing a 2D phase: they respect this symmetry. For this reason, in the islands model we present in Sec.~\ref{sec:phase-control}, we explicitly break the mirror symmetries about all axes.

\bibliography{library}

\end{document}